\documentclass[12pt]{article}
\usepackage{amsmath,amsfonts,amssymb}

\begin{document}
\thispagestyle{empty}

\def\theequation{\arabic{section}.\arabic{equation}}
\def\a{\alpha}
\def\b{\beta}
\def\g{\gamma}
\def\d{\delta}
\def\dd{\rm d}
\def\e{\epsilon}
\def\ve{\varepsilon}
\def\z{\zeta}
\def\B{\mbox{\bf B}}

\newcommand{\h}{\hspace{0.5cm}}

\begin{titlepage}
\renewcommand{\thefootnote}{\fnsymbol{footnote}}
\begin{center}
\Large Finite-size effects of Membranes on $AdS_4\times S_7$
\end{center}
\vskip 1.2cm \centerline{\large Changrim  Ahn and P. Bozhilov
\footnote{On leave from Institute for Nuclear Research and Nuclear
Energy, Bulgarian Academy of Sciences, Bulgaria.}}

\vskip 10mm

\centerline{\sl Department of Physics} \centerline{\sl Ewha Womans
University} \centerline{\sl DaeHyun 11-1, Seoul 120-750, S. Korea}
\vspace*{0.6cm} \centerline{\tt ahn@ewha.ac.kr,
bozhilov@inrne.bas.bg}

\vskip 20mm

\baselineskip 18pt

\begin{center}
{\bf Abstract}
\end{center}
\h We consider semi-classical solutions of membranes on the
$AdS_4\times S^7$ background. This is supposed to be dual to the ${\cal N}=6$
super Chern-Simons theory with $k=1$ in a planar limit recently
proposed by Aharony, Bergmann, Jafferis, and Maldacena (ABJM). We
have identified giant magnon and single spike states on the membrane
by reducing them to the Neumamm - Rosochatius integrable system. We
also connect these to the complex sine-Gordon integrable model.
Based on this approach, we find finite-size membrane solutions and
obtain their images in the complex sine-Gordon system along with the
leading finite-size corrections to the energy-charge relations.
\end{titlepage}
\newpage
\baselineskip 18pt

\def\nn{\nonumber}
\def\tr{{\rm tr}\,}
\def\p{\partial}
\newcommand{\bea}{\begin{eqnarray}}
\newcommand{\eea}{\end{eqnarray}}
\newcommand{\bde}{{\bf e}}
\renewcommand{\thefootnote}{\fnsymbol{footnote}}
\newcommand{\be}{\begin{equation}}
\newcommand{\ee}{\end{equation}}

\vskip 0cm

\renewcommand{\thefootnote}{\arabic{footnote}}
\setcounter{footnote}{0}

\setcounter{equation}{0}
\section{Introduction}

After many interesting developments in the duality between type
IIB string theory on $AdS_5\times S^5$ and ${\cal N}=4$ super
Yang-Mills theory, the AdS/CFT correspondence
\cite{AdS/CFT,GKP98,EW98} is now being extended into the
$AdS_4/CFT_3$. The first three-dimensional conformal field theory,
consistent with all known symmetries of M2-branes was found in
\cite{BL0711}. It is invariant under 16 supersymmetries, possesses
$SO(8)$ $R$-symmetry and is conformally invariant at the classical
level. In this model, the gauge field is nonpropagating. More recent
proposal for the worldvolume theory of multiple M2-branes uses a
Lorentzian three algebra (constructed from ordinary Lie algebra)
\cite{GMR0805,BGTV0805,HIM0805}. There are more related developments
recently \cite{BLS0806,GGRV0806,BKKS,HLLLP}. An alternative proposal
for the theory of multiple M2-branes was made by ABJM which is
${\cal N}=6$ super Chern-Simons theory with $SU(N)\times SU(N)$
gauge symmetry and level $k$ \cite{ABJM0806}. In the limit
$N,k\to\infty$ with a fixed value of 't Hooft coupling
$\lambda=N/k$, this theory is claimed to be dual to the type IIA
superstring theory on $AdS_4\times CP^3$. For a small $\lambda$, a
leading two-loop perturbation calculation has been studied and a new
integrable structure has been discovered \cite{MZ0806,GGY}. This
motivates efforts to discover classical integrability in string
theory side. Indeed, BMN-like states\cite{Taka}, integrability
\cite{AF,Stef}, and giant magnon state and its finite-size effect
\cite{Grignani,Grignanif} in the $k\to\infty$ limit where the target
space becomes $AdS_4\times CP_3$ have been reported.

With these developments, it is interesting to consider dual to
M-theory on $AdS_4\times S^7$ in $\lambda>>1$ limit at $k=1$ by
considering membranes on $AdS_4\times S^7$. It is already known that
there exist M2-brane configurations on $AdS_4\times S^7$, which have
properties, similar to some string solutions on $AdS_5\times S^5$.
In particular, some of them have description in terms of the Neumamm
- Rosochatius (NR) integrable system \cite{MNR}. The NR system has
been proposed for the string theory on $AdS_5\times S^5$ in \cite{AFRT,ART,KRT06}.
It was also
established that they can reproduce the continuous limit of the
integrable $SU(2)$ spin chain \cite{SCM}. (See also \cite{WYW07}.)
Besides, giant magnon (GM) and single spike (SS) like energy-charge
relations have been found \cite{BR06,B06,BR07}. It is interesting if
the above achievements can be extended to include other string
analogies. One possible task is to discover membrane configurations
which can be related to the complex sine-Gordon (CSG) integrable
system. Recently, the finite-size string corrections are actively
investigated as a new window for the AdS/CFT correspondence
\cite{SNZ05,AFZ06,SNZZ06,AFGS,HS0801,MinSax,RamSem,KM0803,AGHO0804,BF0805,AB1}.
That is why, another direction of research can be to try to find
analogous finite-size corrections from M2-branes on $AdS_4\times
S^7$. In this paper, we extend our string results obtained in
\cite{AB1} to the M2-brane case. Namely, we use the reduction of the
M2-brane dynamics to the one of the NR integrable model, to map {\it
all} membrane solutions described by this dynamical system onto
solutions of the CSG integrable model. In the framework of this NR
approach, we find finite size corrections to the membrane
energy-charge relations.

The article is organized as follows. In sect.2 we introduce the
partially gauge fixed M2-brane action and constraints. After
reducing to the $R_t\times S^7$ subspace, we propose membrane
embedding coordinates appropriate for our purposes. In sect.3 we
describe how the NR integrable system arises from the M2-brane. In
sect.4 we find relations between the parameters of the membrane
solutions described by this dynamical system and the parameters in
the corresponding solutions of the CSG integrable model. GM and SS
like solutions are considered as examples. In sect.5 we describe
finite size membrane solution, its image in the complex sine-Gordon
system, and the leading corrections to the energy-charge relations
analogous to the GM and SS strings on $R_t\times S^3$. We conclude
the paper with some comments in Sect.6.

\setcounter{equation}{0}
\section{Membranes on $AdS_4\times S^7$}

We begin with the following membrane action \bea\label{oma} S=\int
d^{3}\xi\left\{\frac{1}{4\lambda^0}\Bigl[G_{00}-2\lambda^{j}G_{0j}+\lambda^{i}
\lambda^{j}G_{ij}-\left(2\lambda^0T_2\right)^2\det G_{ij}\Bigr] +
T_2 C_{012}\right\},\eea where \bea\nn &&G_{mn}= g_{MN}(X)\p_m
X^M\p_n
X^N,\h C_{012}= c_{MNP}(X)\p_{0}X^{M}\p_{1}X^{N}\p_{2}X^{P}, \\
\nn &&\p_m=\p/\p\xi^m,\h m = (0,i) = (0,1,2),\\ \nn
&&(\xi^0,\xi^1,\xi^2)=(\tau,\sigma_1,\sigma_2),\h M =
(0,1,\ldots,10),\eea are the fields induced on the membrane
worldvolume from the background metric $g_{MN}$ and the background
3-form gauge field $c_{MNP}$, $\lambda^m$ are Lagrange
multipliers, $x^M=X^M(\xi)$ are the membrane embedding
coordinates, and $T_2$ is its tension. As shown in \cite{NPB656},
the above action is classically equivalent to the Nambu-Goto type
action \bea\nn S^{NG}= - T_2\int d^{3}\xi
\left(\sqrt{-\det{G_{mn}}}-\frac{1}{6}\varepsilon^{mnp}
\p_{m}X^{M}\p_n X^N \p_{p}X^{P} c_{MNP}\right)\eea and to the
Polyakov type action \bea\nn S^{P}= - \frac{T_2}{2}\int
d^{3}\xi\left[\sqrt{-\gamma}\left(\gamma^{mn} G_{mn}-1\right) -
\frac{1}{3}
\varepsilon^{mnp}\p_{m}X^{M}\p_nX^N\p_{p}X^{P}c_{MNP}\right],\eea
where $\gamma^{mn}$ is the auxiliary worldvolume metric and
$\gamma=\det\gamma_{mn}$. In addition, the action (\ref{oma})
gives a {\it unified} description for the tensile and tensionless
membranes.

The equations of motion for the Lagrange multipliers $\lambda^{m}$
generate the constraints \bea\label{M00}
&&G_{00}-2\lambda^{j}G_{0j}+\lambda^{i}\lambda^{j}G_{ij}
+\left(2\lambda^0T_2\right)^2\det G_{ij}=0,\\
\label{0j} &&G_{0j}-\lambda^{i}G_{ij}=0.\eea

Further on, we will work in diagonal worldvolume gauge
$\lambda^{i}=0$, in which the action (\ref{oma}) and the
constraints (\ref{M00}), (\ref{0j}) simplify to \bea\label{omagf}
&&S_{M}=\int d^{3}\xi \mathcal{L}_{M}= \int
d^{3}\xi\left\{\frac{1}{4\lambda^0}\Bigl[G_{00}-\left(2\lambda^0T_2\right)^2\det
G_{ij}\Bigr] + T_2 C_{012}\right\},
\\ \label{00gf} &&G_{00}+\left(2\lambda^0T_2\right)^2\det G_{ij}=0,
\\ \label{0igf} &&G_{0i}=0.\eea
Let us note that the action (\ref{omagf}) and the constraints
(\ref{00gf}), (\ref{0igf}) {\it coincide} with the usually used
gauge fixed Polyakov type action and constraints after the
following identification of the parameters $2\lambda^0T_2=L=const$
(see for instance \cite{27}).

Searching for membrane configurations in $AdS_4\times S^7$, which
correspond to the Neumann or Neumann-Rosochatius integrable
systems, we should first eliminate the membrane interaction with
the background 3-form field on $AdS_4$, to ensure more close
analogy with the strings on $AdS_5\times S^5$. To make our choice,
let us write down the background. It can be parameterized as
follows \bea\nn &&ds^2=\mathcal{R}^2\left[-\cosh^2\rho
dt^2+d\rho^2+\sinh^2\rho\left(d\alpha^2+\sin^2\alpha
d\beta^2\right)+ 4d\Omega_7^2\right],
\\ \nn &&c_{(3)}=\mathcal{R}^3\sinh^3\rho\sin\alpha dt\wedge
d\alpha\wedge d\beta.\eea

Since we want the membrane to have nonzero conserved energy on
$AdS$, the simplest choice for which the interaction with the
$c_{(3)}$ field disappears, is to fix the coordinates $\rho$,
$\alpha$ and $\beta$: $\rho=0$, $\alpha, \beta=constants$. Thus,
we restrict our considerations to membranes moving on the
$R_t\times S^7$ subspace of $AdS_4\times S^7$ with metric  \bea\nn
&&ds^2_{sub}=\mathcal{R}^2\left\{-dt^2+4\left[d\psi_1^2+\cos^2\psi_1
d\varphi_1^2\right.\right.\\ \nn
&&+\left.\left.\sin^2\psi_1\left(d\psi_2^2+\cos^2\psi_2
d\varphi_2^2+ \sin^2\psi_2\left(d\psi_3^2+\cos^2\psi_3
d\varphi_3^2+\sin^2\psi_3
d\varphi_4^2\right)\right)\right]\right\}.\eea

The membrane embedding into $R_t\times S^7$, appropriate for our
purposes, is \bea\label{mes7} Z_{0}=\mathcal{R}e^{it(\xi^m)}, \h
W_{a}=2\mathcal{R}r_a(\xi^m)e^{i\varphi_a(\xi^m)},\h
a=(1,2,3,4),\eea where $r_a$ are real functions of $\xi^m$, while
$\varphi_a$ are the isometric coordinates on which the background
metric does not depend. The four complex coordinates $W_{a}$ are
restricted by the real embedding condition \bea\label{ecs7}
\delta_{ab}W_{a}\bar{W}_{b}-\left(2\mathcal{R}\right)^2=0,\h
\mbox{or}\h \delta_{ab} r_ar_b-1=0.\eea The coordinates $r_a$ are
connected to the initial coordinates, on which the background
depends, through the equalities \bea\nn &&r_1= \cos\psi_1,\h r_2=
\sin\psi_1\cos\psi_2,
\\ \nn &&r_3= \sin\psi_1\sin\psi_2\cos\psi_3,\h
r_4= \sin\psi_1\sin\psi_2\sin\psi_3.\eea

For the embedding described above, the induced metric is given by
\bea\label{im} &&G_{mn}=-\p_{(m}Z_0\p_{n)}\bar{Z_0} +
\delta_{ab}\p_{(m}W_a\p_{n)}\bar{W_b}= \\ \nn
&&\mathcal{R}^2\left[-\p_mt\p_nt+
4\sum_{a=1}^{4}\left(\p_mr_a\p_nr_a +
r_a^2\p_m\varphi_a\p_n\varphi_a\right)\right].\eea
Correspondingly, the membrane Lagrangian becomes \bea\nn
\mathcal{L}=\mathcal{L}_{M}+\Lambda_M(\delta_{ab} r_ar_b-1),\eea
where $\Lambda_M$ is a Lagrange multiplier.

\setcounter{equation}{0}
\section{NR Integrable System from M2-brane}

Let us consider the following particular case of the above
membrane embedding \cite{MNR} \bea\label{nra}
&&Z_{0}=\mathcal{R}e^{i\kappa\tau},\h
W_{a}=2\mathcal{R}r_a(\xi,\eta)e^{i\left[\omega_{a}\tau+\mu_a(\xi,\eta)\right]},
\\ \nn && \xi=\alpha\sigma_1+\beta\tau,\h \eta=\gamma\sigma_2+\delta\tau,\eea
which implies \bea\label{i} t=\kappa\tau,\h
\varphi_a(\xi^m)=\varphi_a(\tau,\sigma_1,\sigma_2)=
\omega_{a}\tau+\mu_a(\xi,\eta).\eea Here $\kappa$, $\omega_{a}$,
$\alpha$, $\beta$, $\gamma$, $\delta$ are parameters. For this
ansatz, the membrane Lagrangian takes the form ($\p_\xi=\p/\p\xi$,
$\p_\eta=\p/\p\eta$) \bea\nn
&&\mathcal{L}=-\frac{\mathcal{R}^2}{\lambda^0}
\left\{\left(4\lambda^0T_2\mathcal{R}\alpha\gamma\right)^2
\sum_{a<b=1}^{4}\left[(\p_\xi r_a\p_\eta r_b-\p_\eta r_a\p_\xi r_b)^2\right. \right. \\
\nn &&+ \left. \left. (\p_\xi r_a\p_\eta\mu_b-\p_\eta
r_a\p_\xi\mu_b)^2r_b^2 + (\p_\xi\mu_a\p_\eta
r_b-\p_\eta\mu_a\p_\xi r_b)^2r_a^2\right.\right.
\\ \nn &&+\left.\left.(\p_\xi\mu_a\p_\eta\mu_b-\p_\eta\mu_a\p_\xi\mu_b)^2r_a^2r_b^2 \right]\right. \\
\nn &&+
\left.\sum_{a=1}^{4}\left[\left(4\lambda^0T_2\mathcal{R}\alpha\gamma\right)^2
(\p_\xi r_a\p_\eta\mu_a-\p_\eta r_a\p_\xi\mu_a)^2-
\left(\beta\p_\xi\mu_a+\delta\p_\eta\mu_a+\omega_a\right)^2\right]r_a^2\right.
\\
\nn &&-\left.\sum_{a=1}^{4}\left(\beta\p_\xi r_a+\delta\p_\eta
r_a\right)^2+(\kappa/2)^2\right\}+\Lambda_M\left(\sum_{a=1}^{4}r_a^2-1\right).\eea
In order to reduce the above Lagrangian to the NR one, we make the
following choice \bea\nn &&r_{1}=r_{1}(\xi),\h r_{2}=r_{2}(\xi),\h
\omega_3=\pm\omega_4=\omega,\\ \label{NRA}
&&r_3=r_3(\eta)=r_0\sin\eta,\h r_4=r_4(\eta)=r_0\cos\eta,\h r_0<1,\\
\nn &&\mu_1=\mu_1(\xi),\h \mu_2=\mu_2(\xi),\h \mu_3,\mu_4=0,\eea
and receive (prime is used for $d/d\xi$) \bea\nn
&&\mathcal{L}=-\frac{\mathcal{R}^2}{\lambda^0}
\left\{\sum_{a=1}^{2}\left[(\tilde{A}^2-\beta^2)r_a'^2\right.+
\left.(\tilde{A}^2-\beta^2)r_a^2\left(\mu'_a-\frac{\beta\omega_a}{\tilde{A}^2-\beta^2}\right)^2
- \frac{\tilde{A}^2}{\tilde{A}^2-\beta^2}\omega_a^2 r_a^2\right]\right. \\
\nn &&+\left. (\kappa/2)^2-r_0^2(\omega^2+\delta^2)\right\} +
\Lambda_M\left[\sum_{a=1}^{2}r_a^2-(1-r_0^2)\right],\eea where $
\tilde{A}^2\equiv \left(4\lambda^0T_2\mathcal{R} \alpha\gamma
r_0\right)^2$. A single time integration of the equations of
motion for $\mu_a$ following from the above Lagrangian gives
\bea\label{mus}
\mu'_a=\frac{1}{\tilde{A}^2-\beta^2}\left(\frac{C_a}{r_a^2}+\beta\omega_a\right),\eea
where $C_a$ are arbitrary constants. Taking this into account, one
obtains the following effective Lagrangian for the coordinates
$r_a(\xi)$ \bea\nn L_{NR}=
\sum_{a=1}^{2}\left[(\tilde{A}^2-\beta^2)r_a'^2 -
\frac{1}{\tilde{A}^2-\beta^2}\left(\frac{C_a^2}{r_a^2} +
\tilde{A}^2\omega_a^2
r_a^2\right)\right]+\Lambda_M\left[\sum_{a=1}^{2}r_a^2-(1-r_0^2)\right].\eea
This Lagrangian, in full analogy with the string considerations
\cite{KRT06}, corresponds to particular case of the
$n$-dimensional NR integrable system. For $C_a=0$ one obtains the
Neumann integrable system, which in the case at hand describes
two-dimensional harmonic oscillator, constrained to remain on a
circle of radius $\sqrt{1-r_0^2}$.

Let us write down the three constraints (\ref{00gf}), (\ref{0igf})
for the present case. To achieve more close correspondence with
the string on $AdS_5\times S^5$, we want the third one,
$G_{02}=0$, to be identically satisfied. To this end, since
$G_{02}\sim r_0^2\gamma\delta,$ we set $\delta=0$, i.e.
$\eta=\gamma\sigma_2$. Then, the first two constraints give the
conserved Hamiltonian $H_{NR}$ and a relation between the
parameters: \bea\nn
&&H_{NR}=\sum_{a=1}^{2}\left[(\tilde{A}^2-\beta^2)r_a'^2+
\frac{1}{\tilde{A}^2-\beta^2}\left(\frac{C_a^2}{r_a^2} +
\tilde{A}^2\omega_a^2
r_a^2\right)\right]=\frac{\tilde{A}^2+\beta^2}{\tilde{A}^2-\beta^2}
\left[(\kappa/2)^2-(r_0\omega)^2\right],
\\ \label{effcs}&&\sum_{a=1}^{2}\omega_{a}C_a +
\beta\left[(\kappa/2)^2-(r_0\omega)^2\right]=0.\eea For closed
membranes, $r_a$ and $\mu_a$ satisfy the following periodicity
conditions \bea r_a(\xi+2\pi\alpha)=r_a(\xi),\h
\mu_a(\xi+2\pi\alpha)=\mu_a(\xi)+2\pi n_a,\label{pbc} \eea where
$n_a$ are integer winding numbers.

In the case at hand, the background metric does not depend on $t$
and $\varphi_a$. Therefore, the corresponding conserved quantities
are the membrane energy $E$ and four angular momenta $J_a$, given
as spatial integrals of the conjugated to these coordinates
momentum densities \bea\nn E=-\int
d^2\sigma\frac{\p\mathcal{L}}{\p(\p_0 t)},\h J_a=\int
d^2\sigma\frac{\p\mathcal{L}}{\p(\p_0\varphi_a)},\h a=1,2,3,4.\eea
$E$ and $J_a$ can be computed by using the expression (\ref{im})
for the induced metric and the ansats (\ref{nra}), (\ref{i}). In
order to reproduce the string case, we set $\omega=0$, and thus
$J_3=J_4=0$. The energy and the other two angular momenta are
given by \bea\label{cqs}
E=\frac{\pi\mathcal{R}^2\kappa}{\lambda^0\alpha}\int d\xi,\h
J_a=\frac{\pi(2\mathcal{R})^2}{\lambda^0\alpha(\tilde{A}^2-\beta^2)}\int
d\xi \left(\beta C_a + \tilde{A}^2\omega_a r_a^2\right),\h
a=1,2.\eea From here, by using the constraints (\ref{effcs}), one
obtains the energy-charge relation \bea\nn
\frac{4}{\tilde{A}^2-\beta^2}\left[\tilde{A}^2(1-r_0^2) +
\beta\sum_{a=1}^{2}\frac{C_a}{\omega_a}\right]\frac{E}{\kappa}
=\sum_{a=1}^{2}\frac{J_a}{\omega_a}.\eea The corresponding result
for strings on $AdS_5\times S^5$ in conformal gauge is
\cite{KRT06} \bea\nn \frac{1}{\alpha^2-\beta^2}\left(\alpha^2+
\beta\sum_{a}\frac{C_a}{\omega_a}\right)\frac{E}{\kappa}
=\sum_{a}\frac{J_a}{\omega_a}.\eea Obviously, the above membrane
and string energy-charge relations are very similar.

We would like to identically satisfy the embedding condition
\bea\nn \sum_{a=1}^{2}r_a^2-(1-r_0^2)=0.\eea To this end we set
\bea\nn r_1(\xi)=\sqrt{1-r_0^2}\sin\theta(\xi),\h
r_2(\xi)=\sqrt{1-r_0^2}\cos\theta(\xi).\eea Then from
(\ref{effcs}) one finds \bea\label{mtsol} &&\theta'=\frac{\pm
1}{\tilde{A}^2-\beta^2}
\left[(\tilde{A}^2+\beta^2)\tilde{\kappa}^2 -
\frac{\tilde{C}_1^2}{\sin^2{\theta}} -
\frac{\tilde{C}_2^2}{\cos^2{\theta}} -
\tilde{A}^2\left(\omega_1^2\sin^2{\theta}
+\omega_2^2\cos^2{\theta}\right)\right]^{1/2},\\ \nn
&&\sum_{a=1}^{2}\omega_{a}\tilde{C}_a + \beta\tilde{\kappa}^2=0,\h
\tilde{\kappa}^2=\frac{(\kappa/2)^2}{1-r_0^2},\h
\tilde{C}_a^2=\frac{C_a^2}{(1-r_0^2)^2}.\eea By replacing the
solution for $\theta(\xi)$ received from (\ref{mtsol}) into
(\ref{mus}), one obtains the solutions for $\mu_a$:
\bea\label{mu12s}
\mu_1=\frac{1}{\tilde{A}^2-\beta^2}\left(\tilde{C}_1\int\frac{d\xi}{\sin^2{\theta}}
+ \beta\omega_1\xi\right),\h
\mu_2=\frac{1}{\tilde{A}^2-\beta^2}\left(\tilde{C}_2\int\frac{d\xi}{\cos^2{\theta}}
+ \beta\omega_2\xi\right).\eea

\setcounter{equation}{0}
\section{Relationship between NR and CSG Systems}

The CSG system is defined by the Lagrangian \bea\nn
\mathcal{L}(\psi) =
\frac{\eta^{ab}\p_a\bar{\psi}\p_b\psi}{1-\bar{\psi}\psi} +
M^2\bar{\psi}\psi,\h \eta^{ab}=diag(-1,1), \eea which give the
equation of motion \bea\nn \p_a\p^a\psi
+\bar{\psi}\frac{\p_a\psi\p^a\psi}{1-\bar{\psi}\psi} -
M^2(1-\bar{\psi}\psi)\psi=0.\eea If we represent $\psi$ in the
form \bea\nn \psi=\sin(\phi/2)\exp(i\chi/2),\eea the Lagrangian
can be expressed as \bea\nn
\mathcal{L}(\phi,\chi)=\frac{1}{4}\left[\p_a\phi\p^a\phi +
\tan^2(\phi/2)\p_a\chi\p^a\chi + (2M)^2\sin^2(\phi/2)\right],\eea
while the equations of motion take the form \bea\label{fem}
&&\p_a\p^a\phi -
\frac{1}{2}\frac{\sin(\phi/2)}{\cos^3(\phi/2)}\p_a\chi\p^a\chi -
M^2\sin\phi=0,\\ \label{kem} &&\p_a\p^a\chi +
\frac{2}{\sin\phi}\p_a\phi\p^a\chi=0.\eea The SG system
corresponds to a particular case of $\chi=0$.

To relate the NR with CSG integrable system, we consider the case
\bea\nn \phi=\phi(\xi),\h \chi=A\sigma_1+B\tau +
\tilde{\chi}(\xi), \eea where $\phi$ and $\tilde{\chi}$ depend on
only one variable $\xi=\alpha\sigma_1+\beta\tau$ in the same way
as in our NR ansatz (\ref{NRA}). Then the equations of motion
(\ref{fem}), (\ref{kem}) reduce to \bea\label{fer} &&\phi'' -
\frac{1}{2}\frac{\sin(\phi/2)}{\cos^3(\phi/2)}
\left[\tilde{\chi}'^2 +
2\frac{A\alpha-B\beta}{\alpha^2-\beta^2}\tilde{\chi}' +
\frac{A^2-B^2}{\alpha^2-\beta^2}\right]
- \frac{M^2\sin\phi}{\alpha^2-\beta^2}=0,\\
\label{ker} &&\tilde{\chi}'' +
\frac{2\phi'}{\sin\phi}\left(\tilde{\chi}' +
\frac{A\alpha-B\beta}{\alpha^2-\beta^2}\right)=0.\eea

We further restrict ourselves to the case of $A\alpha=B\beta$. A
trivial solution of Eq.(\ref{ker}) is $\tilde{\chi}=constant$,
which corresponds to the solutions of the CSG equations considered
in \cite{CDO06,OS06} for a GM string on $R_t\times S^3$. More
nontrivial solution of (\ref{ker}) is \bea\label{kfi}
\tilde{\chi}= C_\chi\int \frac{d\xi}{\tan^2(\phi/2)}.\eea The
replacement of the above into (\ref{fer}) gives
\bea\label{fef}\phi''=\frac{M^2\sin\phi}{\alpha^2-\beta^2} +
\frac{1}{2}\left[C_\chi^2\frac{\cos(\phi/2)}{\sin^3(\phi/2)}
-\frac{A^2}{\beta^2}\frac{\sin(\phi/2)}{\cos^3(\phi/2)}\right].\eea
Integrating once, we obtain \bea\label{ffi}
\phi'&=&\pm\left[\left(C_\phi -
\frac{2M^2}{\alpha^2-\beta^2}\right) +
\frac{4M^2}{\alpha^2-\beta^2}\sin^2(\phi/2)
-\frac{A^2/\beta^2}{1-\sin^2(\phi/2)} -
\frac{C_\chi^2}{\sin^2(\phi/2)}\right]^{1/2}\\
&\equiv&\pm\Phi(\phi),\nonumber \eea from which we get \bea\nn
\xi(\phi)=\pm\int \frac{d\phi}{\Phi(\phi)},\qquad\chi(\phi)=
\frac{A}{\beta}\left(\beta\sigma+\alpha\tau\right)\pm C_\chi\int
\frac{d\phi}{\tan^2(\phi/2)\Phi(\phi)}.\eea All these solve the
CSG system for the considered particular case. It is clear from
(\ref{ffi}) that the expression inside the square root must be
positive.

After the reduction of our membrane configuration to a NR-type
integrable system, we can establish relationship between its
solutions and the solutions of the reduced CSG system, as
described above. With this aim, we make the following
identification \bea\label{Mequiv} \sin^2(\phi/2)\equiv
\frac{\sqrt{-G}}{\tilde{K}^2},\eea where $G$ is the determinant of
the metric $G_{mn}$ induced on the membrane worldvolume, computed
on the constraints (\ref{00gf}), (\ref{0igf}), and $\tilde{K}^2$
is a parameter. For the NR system obtained from the M2-brane,
$\sqrt{-G}$ is given by \bea\label{det} \sqrt{-G}
=Q^2\frac{\mathcal{R}^2\tilde{A}^2}{\tilde{A}^2-\beta^2}
\left[(\tilde{\kappa}^2-\omega_1^2) +
(\omega_1^2-\omega_2^2)\cos^2\theta\right],\h
Q^2=\frac{1-r_0^2}{(\lambda^0T_2)^2}.\eea To relate the parameters
in the solutions of the NR and CSG integrable systems, we put
(\ref{Mequiv}), (\ref{det}) into (\ref{ffi}) and receive \bea\nn
&&\tilde{K}^2=(Q\mathcal{R}M)^2
\left(\frac{1-\tilde{A}^2/\beta^2}{1-\alpha^2/\beta^2}\right)
\equiv (Q\mathcal{R})^2\tilde{M}^2,\\
\nn &&C_\phi= \frac{2}{\tilde{A}^2-\beta^2}\left[3\tilde{M}^2- 2
\left(\tilde{\kappa}^2 + Y -\Omega\right)\right], \eea
\bea\label{prels}
&&\frac{1}{4}\tilde{M}^4(\tilde{A}^2-\beta^2)\frac{A^2}{\beta^2}=
\tilde{M}^4\left(\tilde{M}^2-\tilde{\kappa}^2 +\Omega\right)
\\ \nn &&-Y
\left[\tilde{M}^4 + \left(\tilde{M}^2- Y\right)
Y-\left(2\tilde{M}^2- Y\right)
\left(\tilde{\kappa}^2-\Omega\right)\right]
\\ \nn &&-\frac{(\omega_1^2-\omega_2^2)}{\omega_1^2\left(1-\frac{\beta^2}{\tilde{A}^2}\right)^3}
\left\{\left[\tilde{M}^2\left(1-\frac{\beta^2}{\tilde{A}^2}\right)
- \tilde{\kappa}^2\right]
(\omega_1^2-\omega_2^2)\hat{C}_2^2\right.\\ \nn
&&-\left.\left[\tilde{M}^2\left(1-\frac{\beta^2}{\tilde{A}^2}\right)
- (\tilde{\kappa}^2-\omega_1^2)\right]
\left[2\frac{\beta}{\tilde{A}}\omega_2\tilde{\kappa}^2\hat{C}_2 +
\left(\tilde{\kappa}^2-\omega_1^2\right)
\left(\frac{\beta^2}{\tilde{A}^2}\tilde{\kappa}^2-\omega_1^2\right)\right]\right\},\eea
\bea \nn &&\frac{1}{4}\tilde{M}^4(\tilde{A}^2-\beta^2) C_\chi^2=
Y^2 \left(Y +\Omega-\tilde{\kappa}^2\right)
\\ \nn &&+\frac{(\omega_1^2-\omega_2^2)}{\omega_1^2
\left(1-\frac{\beta^2}{\tilde{A}^2}\right)^3}
\left\{\tilde{\kappa}^2
(\omega_1^2-\omega_2^2)\hat{C}_2^2-(\tilde{\kappa}^2-\omega_1^2)
\left[2\frac{\beta}{\tilde{A}}\omega_2\tilde{\kappa}^2\hat{C}_2 +
\left(\tilde{\kappa}^2-\omega_1^2\right)
\left(\frac{\beta^2}{\tilde{A}^2}\tilde{\kappa}^2-\omega_1^2\right)\right]\right\},\eea
where \bea\nn Y=
\frac{\tilde{\kappa}^2-\omega_1^2}{1-\frac{\beta^2}{\tilde{A}^2}},\h
\Omega= \frac{\omega_2^2}
{1-\frac{\beta^2}{\tilde{A}^2}},\h\hat{C}_2=\tilde{C}_2/\tilde{A}.\eea
In this way, we expressed the CSG parameters $C_\phi$, $A$ and
$C_\chi$ through the NR parameters $\tilde{A}$, $\beta$,
$\tilde{\kappa}$, $\omega_1$, $\omega_2$, $\tilde{C}_2$. The mass
parameter $\tilde{M}^2$ remains free. The above equalities give
the mapping, which relates all membrane solutions derivable from
(\ref{mtsol}), (\ref{mu12s}) with the corresponding solutions of
the CSG system.

\subsection{Examples: GM and SS analogues}

In particular, for the GM-type membrane solutions, for which
$\tilde{C}_2=0$, $\tilde{\kappa}^2=\omega_1^2$, the following
equalities between the parameters hold \bea\nn &&C_\phi=
\frac{2}{\tilde{A}^2-\beta^2}\left[3\tilde{M}^2- 2\left(\omega_1^2
-\frac{\omega_2^2}{1-\beta^2/\tilde{A}^2}\right)\right],\h
\tilde{K}^2=(Q\mathcal{R})^2\tilde{M}^2,
\\ \label{MGMf} &&A^2=\frac{4}{\tilde{A}^2/\beta^2-1}\left(\tilde{M}^2-\omega_1^2
+ \frac{\omega_2^2}{1-\beta^2/\tilde{A}^2}\right), \h
C_{\chi}=0.\eea For the case of SS-type membrane solutions, when
$\tilde{C}_2=0$, $\tilde{\kappa}^2=\omega_1^2\tilde{A}^2/\beta^2$,
one has \bea\nn &&C_\phi= \frac{2}{\beta^2-\tilde{A}^2}\left[
2\left(2\omega_1^2\tilde{A}^2/\beta^2
+\frac{\omega_2^2}{\beta^2/\tilde{A}^2-1}\right)-3\tilde{M}^2\right], \\
\label{MSSf} &&A^2=\frac{4}{\tilde{M}^4(1-\tilde{A}^2/\beta^2)}
\left(\omega_1^2\tilde{A}^2/\beta^2-\tilde{M}^2\right)^2
\left(\frac{\omega_2^2}{\beta^2/\tilde{A}^2-1}-\tilde{M}^2\right),
\\ \nn &&C_{\chi}=\frac{2\omega_1^2\omega_2\tilde{A}^3}
{\tilde{M}^2(\beta^2-\tilde{A}^2)\beta^2},\h
\tilde{K}^2=(Q\mathcal{R})^2\tilde{M}^2.\eea

Let us give the membrane configurations which are analogous to the
GM and SS string solutions on $R_t\times S^3$.

For the GM-like case by using that  $\tilde{C}_2=0$,
$\tilde{\kappa}^2=\omega_1^2$ in (\ref{mtsol}), (\ref{mu12s}), one
finds \bea\nn
&&\cos\theta=\frac{\cos\tilde{\theta}_0}{\cosh\left(D_0\xi\right)},\h
\sin^2\tilde{\theta}_0=\frac{\beta^2\omega_1^2}{\tilde{A}^2(\omega_1^2-\omega_2^2)},\h
D_0=\frac{\tilde{A}\sqrt{\omega_1^2-\omega_2^2}}
{\tilde{A}^2-\beta^2}\cos\tilde{\theta}_0,
\\ \nn &&\mu_1=\arctan\left[\cot\tilde{\theta}_0\tanh(D_0\xi)\right],
\h\mu_2=\frac{\beta\omega_2}{\tilde{A}^2-\beta^2}\xi .\eea Then,
the corresponding membrane configuration is given by \bea\nn
&&Z_0=\mathcal{R}\exp\left(2i\sqrt{1-r_0^2}\omega_1\tau\right), \\
\nn &&W_1=2\mathcal{R}\sqrt{1-r_0^2}
\sqrt{1-\frac{\cos^2\tilde{\theta}_0}{\cosh^2\left(D_0\xi\right)}}
\exp\left\{i\omega_1\tau +
i\arctan\left[\cot\tilde{\theta}_0\tanh(D_0\xi)\right]\right\},
\\ \label{MGM} &&W_2=2\mathcal{R}\sqrt{1-r_0^2}
\frac{\cos\tilde{\theta}_0}{\cosh\left(D_0\xi\right)}
\exp\left[i\omega_2\left(\tau +
\frac{\beta}{\tilde{A}^2-\beta^2}\xi\right)\right],
\\ \nn &&W_3=2\mathcal{R}r_0\sin(\gamma\sigma_2),
\\ \nn &&W_4=2\mathcal{R}r_0\cos(\gamma\sigma_2).\eea

For the SS-like solutions when $\tilde{C}_2=0$,
$\tilde{\kappa}^2=\omega_1^2\tilde{A}^2/\beta^2$, by solving the
equations (\ref{mtsol}), (\ref{mu12s}), one arrives at \bea\nn
&&\cos\theta=\frac{\cos\tilde{\theta}_1}{\cosh\left(D_1\xi\right)},\h
\sin^2\tilde{\theta}_1=\frac{\tilde{A}^2\omega_1^2}{\beta^2(\omega_1^2-\omega_2^2)},\h
D_1=\frac{\tilde{A}\sqrt{\omega_1^2-\omega_2^2}}
{\tilde{A}^2-\beta^2}\cos\tilde{\theta}_1,
\\ \nn &&\mu_1= -\frac{\omega_1}{\beta}\xi
-\arctan\left[\cot\tilde{\theta}_1\tanh(D_1\xi)\right],
\h\mu_2=\frac{\beta\omega_2}{\tilde{A}^2-\beta^2}\xi .\eea Now,
the shape of the membrane is described by \bea\nn
&&Z_0=\mathcal{R}\exp\left(2i\sqrt{1-r_0^2}
\frac{\tilde{A}}{\beta}\omega_1\tau\right), \\
\nn &&W_1=2\mathcal{R}\sqrt{1-r_0^2}
\sqrt{1-\frac{\cos^2\tilde{\theta}_1}{\cosh^2\left(D_1\xi\right)}}
\exp\left\{-i\omega_1\frac{\alpha}{\beta}\sigma_1 -
i\arctan\left[\cot\tilde{\theta}_1\tanh(D_1\xi)\right]\right\},
\\ \label{MSS} &&W_2=2\mathcal{R}\sqrt{1-r_0^2}
\frac{\cos\tilde{\theta}_1}{\cosh\left(D_1\xi\right)}
\exp\left[i\omega_2\left(\tau +
\frac{\beta}{\tilde{A}^2-\beta^2}\xi\right)\right],
\\ \nn &&W_3=2\mathcal{R}r_0\sin(\gamma\sigma_2),
\\ \nn &&W_4=2\mathcal{R}r_0\cos(\gamma\sigma_2).\eea

The energy-charge relations computed on the above membrane
solutions were found in \cite{BR07}, and in our notations read
\bea\label{EJGM0} \sqrt{1-r_0^2}E-\frac{J_1}{2}
=\sqrt{\left(\frac{J_2}{2}\right)^2
+\frac{\tilde{\lambda}}{\pi^2}\sin^2\frac{p}{2}},\h
\frac{p}{2}=\frac{\pi}{2}-\tilde{\theta}_0, \eea for the GM-like
case, and \bea\label{EJSS0}
\sqrt{1-r_0^2}E-\frac{\sqrt{\tilde{\lambda}}}{2\pi}\Delta\varphi_1
=\frac{\sqrt{\tilde{\lambda}}}{\pi}\frac{p}{2}, \h \frac{J_1}{2}
=\sqrt{\left(\frac{J_2}{2}\right)^2
+\frac{\tilde{\lambda}}{\pi^2}\sin^2\frac{p}{2}},\h
\frac{p}{2}=\frac{\pi}{2}-\tilde{\theta}_1, \eea for the SS-like
solution, where \bea\label{tl}
\tilde{\lambda}=\left[(4\pi)^2T_2\mathcal{R}^3r_0
(1-r_0^2)\gamma\right]^2.\eea

For the obtained membrane solutions (\ref{MGM}), (\ref{MSS}), one
has \bea\nn |W_1|^2+|W_2|^2=(2\mathcal{R})^2(1-r_0^2),\h
W_3^2+W_4^2=(2\mathcal{R}r_0)^2.\eea That is why, these membrane
configurations live in the $R_t\times S^3\times S^1$ subspace of
$AdS_4\times S^7$. Besides, the radii of the three-sphere $S^3$
and the circle $S^1$ are functions of the parameter $r_0$. When
$r_0$ approaches 1 from below, the radius of $S^3$ decreases,
while the radius of $S^1$ increases. For $r_0 \to 0$, we observe
the opposite behavior.

One may ask what happens when $r_0 = 0$. As is seen from
(\ref{MGM}) and (\ref{MSS}), the membrane shrinks to a string in
this case, because the dependence on $\sigma_2$ disappears.
However, this string has completely different properties. Indeed,
by solving the Eqs. (\ref{mtsol}), (\ref{mu12s}), one finds the
following string solution (now $\tilde{A}=0$) \bea\nn
&&Z_0=\mathcal{R}\exp\left(2i\tilde{\kappa}\tau\right), \\
\nn &&W_1=2\mathcal{R}
\sqrt{1-\left(1-\frac{\tilde{\kappa}^2}{\omega_1^2}\right)
\sin^2\left(\frac{\tilde{\kappa}}{\beta}\xi\right)}
\exp\left[i\omega_1\tau + i\frac{\tilde{\kappa}}{\omega_1}
F\left(\frac{\tilde{\kappa}}{\beta}\xi \bigg \vert
1-\frac{\tilde{\kappa}^2}{\omega_1^2}\right)\right],
\\ \nn &&W_2=2\mathcal{R}\sqrt{1-\frac{\tilde{\kappa}^2}{\omega_1^2}
}\sin\left(\frac{\tilde{\kappa}}{\beta}\xi\right)
\exp\left(-i\omega_2 \frac{\alpha}{\beta}\sigma_1\right),\eea
where $F(n|m)$ is the elliptic integral of first kind. The
computation of the corresponding conserved quantities gives
\bea\nn E-\frac{J_1}{2}=\frac{(2\pi\mathcal{R})^2}{\lambda^0}
\tilde{\kappa}\left(1-\frac{\tilde{\kappa}}{\omega_1}\right), \h
J_2=0.\eea If we set here $\tilde{\kappa}=\omega_1$ as for the GM
case, we obtain $E-J_1/2=0$, i.e. the vacuum state. The same
result can be obtained directly from (\ref{EJGM0}), taking into
account (\ref{cqs}), since for $r_0 = 0$ we have
$\tilde{\lambda}=0$ too. The SS case corresponds to
$\tilde{\kappa}=0$, which leads to trivial solution with
$E=J_1=J_2=0$.

Let us explain the obvious differences between the M2-brane
energy-charge relation (\ref{EJGM0}) and the one for dyonic GM
strings on $R_t\times S^3$, which as is well known is given by
\bea\nn E-J_1=\sqrt{J_2^2 + \frac{\lambda}{\pi^2}
\sin^2\frac{p}{2}}.\eea The factor $ \sqrt{1-r_0^2}$ comes from
the fact that the NR system for membranes is defined to live on a
circle with radius $ \sqrt{1-r_0^2}$, while for the strings this
radius is one. The factor $1/2$ appears as a consequence of the
background geometry. While the radii of $AdS_5$ and $S^5$ in the
type IIB background $AdS_5\times S^5$ are equal, the radius of
$AdS_4$ is half the $S^7$ radius in the $AdS_4\times S^7$ target
space. The same applies to the SS case. Note however that such
coefficients in the dispersion relation can also appear for
strings on $AdS_5\times S^5$ as shown in \cite{BR06,B06}.

Let us also write down the images of these M2-brane solutions in
the CSG system. In order to derive them, we replace (\ref{MGMf})
for the GM-like case and correspondingly (\ref{MSSf}) for the
SS-like case into (\ref{ffi}), and then use the obtained field
$\phi$ in (\ref{kfi}) in order to find $\chi$. The results of the
integrations are as follows: \bea\nn \Psi_{GM-like}&=&
\frac{\sqrt{\omega_1^2-\omega_2^2/(1-\beta^2/\tilde{A}^2)}}
{\tilde{M}
\cosh\left[\sqrt{\frac{\omega_1^2-\omega_2^2/(1-\beta^2/\tilde{A}^2)}
{1-\beta^2/\tilde{A}^2}}(\xi/\tilde{A})\right]}\\ \nn &&\times
\exp\left[i\sqrt{\frac{\tilde{M}^2-\omega_1^2+
\omega_2^2/(1-\beta^2/\tilde{A}^2)}
{\tilde{A}^2/\beta^2-1}}\left(\sigma_1+
\frac{\alpha}{\beta}\tau\right)\right],\eea \bea\nn
\Psi_{SS-like}&=&\sqrt{\tanh^2\left(D\xi\right) +
\frac{\omega_2^2}{\omega_1^2\left(1-\tilde{A}^2/\beta^2\right)\cosh^2
\left(D\xi\right)}}\\
\nn &&\times \exp\left\{i\arctan\left[\frac{\omega_1}{\omega_2}
\sqrt{1-\tilde{A}^2/\beta^2-\omega_2^2/\omega_1^2}
\tanh\left(D\xi\right)\right]\right\},\eea where \bea\nn
D=\frac{\tilde{A}\omega_1\sqrt{1-\tilde{A}^2/\beta^2-\omega_2^2/\omega_1^2}}
{\beta^2\left(1-\tilde{A}^2/\beta^2\right)}.\eea We point out that
the obtained $\Psi_{SS-like}$ corresponds to
$\tilde{M}^2=\tilde{\kappa}^2=\omega_1^2\tilde{A}^2/\beta^2$, when
the parameter $A$ in (\ref{MSSf}) becomes zero.

\setcounter{equation}{0}
\section{Finite-Size Effects}

In this section we will find finite-size membrane solution, its
image in the CSG system, and the leading corrections to the
energy-charge relations analogous to the ones for the GM and SS
strings on $R_t\times S^3$.

For $C_2=0$, Eq.(\ref{mtsol}) can be written as \bea\label{tS3eq}
(\cos\theta)'=\mp\frac{\tilde{A}\sqrt{\omega_1^2-\omega_2^2}}{\tilde{A}^2-\beta^2}
\sqrt{(z_+^2-\cos^2\theta)(\cos^2\theta-z_-^2)},\eea where \bea\nn
&&z^2_\pm=\frac{1}{2(1-\frac{\omega_2^2}{\omega_1^2})}
\left\{q_1+q_2-\frac{\omega_2^2}{\omega_1^2}
\pm\sqrt{(q_1-q_2)^2-\left[2\left(q_1+q_2-2q_1
q_2\right)-\frac{\omega_2^2}{\omega_1^2}\right]
\frac{\omega_2^2}{\omega_1^2}}\right\}, \\ \nn
&&q_1=1-\tilde{\kappa}^2/\omega_1^2,\h
q_2=1-\beta^2\tilde{\kappa}^2/\tilde{A}^2\omega_1^2 .\eea The
solution of (\ref{tS3eq}) is \bea\label{tS3sol} \cos\theta=z_+
dn\left(C\xi|m\right),\h
C=\mp\frac{\tilde{A}\sqrt{\omega_1^2-\omega_2^2}}{\tilde{A}^2-\beta^2}
z_+,\h m\equiv 1-z^2_-/z^2_+ .\eea The solutions of
Eqs.(\ref{mu12s}) now read \bea\nn
&&\mu_1=\frac{2\beta/\tilde{A}}{z_+\sqrt{1-\omega_2^2/\omega_1^2}}
\left[F\left(am(C\xi)|m\right) -
\frac{\tilde{\kappa}^2/\omega_1^2}{1-z^2_+}\Pi\left(am(C\xi),-\frac{z^2_+
-z^2_-}{1-z^2_+}\bigg \vert m\right)\right],\\ \nn
&&\mu_2=\frac{2\beta\omega_2/\tilde{A}\omega_1}{z_+\sqrt{1-\omega_2^2/\omega_1^2}}
F\left(am(C\xi)|m\right),\eea where $\Pi(k,n|m)$ is the elliptic
integral of third kind. Therefore, the full membrane solution
is given by \bea\nn &&Z_0=\mathcal{R}\exp(2i\sqrt{1-r_0^2}\tilde{\kappa}\tau),\\
\nn &&W_1=2\mathcal{R}\sqrt{1-r_0^2}\sqrt{1-z^2_+
dn^2\left(C\xi|m\right)}\exp\left\{i\omega_1\tau +
\frac{2i\beta/\tilde{A}}{z_+\sqrt{1-\omega_2^2/\omega_1^2}}\right.\\
\nn &&\times\left.\left[F\left(am(C\xi)|m\right) -
\frac{\tilde{\kappa}^2/\omega_1^2}{1-z^2_+}\Pi\left(am(C\xi),-\frac{z^2_+
-z^2_-}{1-z^2_+}\bigg \vert m\right)\right]\right\} ,\\
\label{fssS3} &&W_2=2\mathcal{R}\sqrt{1-r_0^2}z_+
dn\left(C\xi|m\right)\exp\left[i\omega_2\tau
+\frac{2i\beta\omega_2/\tilde{A}\omega_1}
{z_+\sqrt{1-\omega_2^2/\omega_1^2}}
F\left(am(C\xi)|m\right)\right],
\\ \nn &&W_3=2\mathcal{R}r_0\sin(\gamma\sigma_2),
\\ \nn &&W_4=2\mathcal{R}r_0\cos(\gamma\sigma_2) .\eea
We note that (\ref{fssS3}) contains both cases:
$\tilde{A}^2>\beta^2$ and $\tilde{A}^2<\beta^2$ corresponding to
GM and SS respectively.

To find the CSG solution related to (\ref{fssS3}), we insert
(\ref{tS3sol}) into (\ref{Mequiv}) and (\ref{det}) to get
\bea\label{fisS3}
\sin^2(\phi/2)=\frac{\omega_1^2/\tilde{M}^2}{\beta^2/\tilde{A}^2-1}
\left[\left(1-\tilde{\kappa}^2/\omega_1^2\right) -
\left(1-\omega_2^2/\omega_1^2\right)\left(z^2_+ cn^2(C\xi|m) +
z^2_- sn^2(C\xi|m)\right)\right].\eea After that, we put
(\ref{fisS3}) into (\ref{kfi}) and integrate. The result is as
follows \bea\label{chiS3}
\chi=\frac{A}{\beta}(\beta\sigma_1+\alpha\tau) -
C_\chi(\alpha\sigma_1+\beta\tau) + \frac{C_\chi}{C
D}\Pi\left(am(C\xi),n|m\right),\eea where $A/\beta$ and $C_\chi$
are given in (\ref{prels}) ($\hat{C}_2=0$), and \bea\nn
D=\frac{\omega_1^2/\tilde{M}^2}{\beta^2/\tilde{A}^2-1}
\left[\left(1-\tilde{\kappa}^2/\omega_1^2\right) -
\left(1-\omega_2^2/\omega_1^2\right)z^2_+\right],\h
n=\frac{\left(1-\omega_2^2/\omega_1^2\right)
(z^2_+-z^2_-)}{\left(1-\tilde{\kappa}^2/\omega_1^2\right) -
\left(1-\omega_2^2/\omega_1^2\right)z^2_+}.\eea Hence for the
present case, the CSG field $\psi=\sin(\phi/2)\exp(i\chi/2)$ is
determined by (\ref{fisS3}) and (\ref{chiS3}).

Our next task is to find out what kind of energy-charge relations
can appear for the M2-brane solution (\ref{fssS3}) in the limit
when the energy $E\to \infty$. It turns out that the semiclassical
behavior depends crucially on the sign of the difference
$\tilde{A}^2-\beta^2$.

\subsection{The GM analogue}

We begin with the GM analogue, i.e. $\tilde{A}^2>\beta^2$. In this
case, one obtains from (\ref{cqs}) the following expressions for
the conserved energy $E$ and the angular momenta $J_1$, $J_2$
\bea\nn &&\mathcal{E}
=\frac{2\tilde{\kappa}(1-\beta^2/\tilde{A}^2)} {\omega_1
z_+\sqrt{1-\omega_2^2/\omega_1^2}}\mathbf{K}
\left(1-z^2_-/z^2_+\right), \\ \label{cqsGM} &&\mathcal{J}_1=
\frac{2 z_+}{\sqrt{1-\omega_2^2/\omega_1^2}} \left[
\frac{1-\beta^2\tilde{\kappa}^2/\tilde{A}^2\omega_1^2}{z^2_+}\mathbf{K}
\left(1-z^2_-/z^2_+\right)-\mathbf{E}
\left(1-z^2_-/z^2_+\right)\right], \\ \nn &&\mathcal{J}_2= \frac{2
z_+ \omega_2/\omega_1 }{\sqrt{1-\omega_2^2/\omega_1^2}}\mathbf{E}
\left(1-z^2_-/z^2_+\right).\eea Here, we have used the notations
\bea\label{not} \mathcal{E}=\frac{2\pi}{\sqrt{\tilde{\lambda}}}
\sqrt{1-r_0^2} E ,\h
\mathcal{J}_1=\frac{2\pi}{\sqrt{\tilde{\lambda}}}\frac{J_1}{2}, \h
\mathcal{J}_2=\frac{2\pi}{\sqrt{\tilde{\lambda}}}\frac{J_2}{2},\eea
where $\tilde{\lambda}$ is defined in (\ref{tl}). The computation
of $\Delta\varphi_1$ gives \bea\label{pws} p\equiv\Delta\varphi_1
&=& 2\int_{\theta_{min}}^{\theta_{max}}\frac{d
\theta}{\theta'}\mu'_1=
\\ \nn &-&\frac{2\beta/\tilde{A}}{z_+\sqrt{1-\omega_2^2/\omega_1^2}}
\left[\frac{\tilde{\kappa}^2/\omega_1^2}{1-z^2_+}\Pi\left(-\frac{z^2_+
- z^2_-}{1-z^2_+}\bigg\vert 1-z^2_-/z^2_+\right) -\mathbf{K}
\left(1-z^2_-/z^2_+\right)\right].\eea In the above expressions,
$\mathbf{K}(m)$, $\mathbf{E}(m)$ and $\Pi(n|m)$ are the complete
elliptic integrals.

Let us introduce the new parameters \bea\nn
u\equiv\omega^2_2/\omega^2_1,\h v\equiv -\beta/\tilde{A},\h
\epsilon\equiv z^2_-/z^2_+ .\eea This will allow us to eliminate
$\tilde{\kappa}/\omega_1$ and $z_\pm$ from the coefficients in
(\ref{cqsGM}), (\ref{pws}) and rewrite them as functions of $u$,
$v$ and $\epsilon$ only: \bea\nn
&&\mathcal{E}=2 K_e \mathbf{K}\left(1-\epsilon\right), \\
\label{ncoeffGM} &&\mathcal{J}_1=2 K_{11}\left[
K_{12}\mathbf{K}\left(1-\epsilon\right)-\mathbf{E}\left(1-\epsilon\right)\right],
\\ \nn &&\mathcal{J}_2=2 K_2 \mathbf{E}\left(1-\epsilon\right), \\
\nn &&p=2 K_{\varphi 1} \left[K_{\varphi 2}\Pi\left(K_{\varphi
3}|1-\epsilon\right) -\mathbf{K}
\left(1-\epsilon\right)\right].\eea We are interested in the
behavior of these quantities in the limit $\epsilon\to 0$. To
establish it, we will use the expansions for the elliptic
integrals and $K_e,\ldots,K_{\varphi 3}$ given in appendix A.

Our approach is as follows. First, we expand $\mathcal{E}$,
$\mathcal{J}_1$, $\mathcal{J}_2$ and $p$ about $\epsilon=0$
keeping $u$ and $v$ independent of $\epsilon$. Second, we
introduce $u(\epsilon)$ and $v(\epsilon)$ according to the rule
\bea\label{uvexp} u(\epsilon)=u_0+u_1\epsilon +
u_2\epsilon\log(\epsilon),\h  v(\epsilon)=v_0+v_1\epsilon +
v_2\epsilon\log(\epsilon)\eea and expand again. Requiring
$\mathcal{J}_2$ and $p$ to be finite, we find \bea\nn
&&u_0=\frac{\mathcal{J}_2^2}{\mathcal{J}_2^2+4\sin^2(p/2)},\h
v_0=\frac{\sin(p)}{\sqrt{\mathcal{J}_2^2+4\sin^2(p/2)}} \\
\nn &&u_1=\frac{u_0}{2(1-v_0^2)}
\left\{1-4v_0^2+3v_0^4-(1-v_0^2)^2\log(16)\right.\\ \nn &&\h\h-u_0
\left.\left[1-\log(16)+3v_0^2\left(\log(16)-3\right)\right]\right\}
\\ \nn &&v_1=\frac{v_0}{4(1-u_0)(1-v_0^2)} \left\{u_0^2(1+3v_0^2)
\left(\log(16)-3\right) +
(1-v_0^2)^2\left(\log(16)-1\right)\right.
\\ \nn &&\h\h-u_0(1-v_0^2)
\left.\left[v_0^2\left(\log(16)-3\right)+\log(256)-4\right]\right\}
\\ \nn &&u_2=\frac{u_0}{2(1-v_0^2)} \left[(1-v_0^2)^2-u_0(1-3v_0^2)\right]
\\ \nn &&v_2=-\frac{1}{4}v_0 \left[u_0 +
\frac{(1-2u_0-v_0^2)\left(1-u_0 - (1+u_0)v_0^2\right)}
{(1-u_0)(1-v_0^2)}\right].\eea The parameter $\epsilon$ can be
obtained from the expansion for $\mathcal{J}_1$ to be \bea\nn
\epsilon=16 \exp\left[-\frac{\sqrt{1-u_0-v_0^2}\mathcal{J}_1 +
2\left(1-v_0^2/(1-u_0)\right)}{1-v_0^2}\right].\eea Using all of
the above in the expansion for $\mathcal{E}-\mathcal{J}_1$, one
arrives at \bea\label{IEJ1} &&\mathcal{E}-\mathcal{J}_1 =
\sqrt{\mathcal{J}_2^2+4\sin^2(p/2)} - \frac{16 \sin^4(p/2)}
{\sqrt{\mathcal{J}_2^2+4\sin^2(p/2)}}\\ \nn
&&\exp\left[-\frac{2\left(\mathcal{J}_1 +
\sqrt{\mathcal{J}_2^2+4\sin^2(p/2)}\right)
\sqrt{\mathcal{J}_2^2+4\sin^2(p/2)}\sin^2(p/2)}{\mathcal{J}_2^2+4\sin^4(p/2)}
\right].\eea It is easy to check that the energy-charge relation
(\ref{IEJ1}) coincides with the one found in \cite{HS08},
describing the finite-size effects for dyonic GM. The difference
is that in the string case the relations between $\mathcal{E}$,
$\mathcal{J}_1$, $\mathcal{J}_2$ and $E$, $J_1$, $J_2$ are given
by \bea\nn \mathcal{E}=\frac{2\pi}{\sqrt{\lambda}}E ,\h
\mathcal{J}_1=\frac{2\pi}{\sqrt{\lambda}}J_1, \h
\mathcal{J}_2=\frac{2\pi}{\sqrt{\lambda}}J_2,\eea while for the
M2-brane they are written in (\ref{not}).

\subsection{The SS analogue}

Let us turn our attention to the SS analogue, when
$\tilde{A}^2<\beta^2$. The computation of the conserved quantities
(\ref{cqs}) and $\Delta\varphi_1$ now gives \bea\nn &&\mathcal{E}
=\frac{2\tilde{\kappa}(\beta^2/\tilde{A}^2-1)}
{\omega_1\sqrt{1-\omega_2^2/\omega_1^2}z_+}\mathbf{K}
\left(1-z^2_-/z^2_+\right), \\ \nn &&\mathcal{J}_1= \frac{2
z_+}{\sqrt{1-\omega_2^2/\omega_1^2}} \left[\mathbf{E}
\left(1-z^2_-/z^2_+\right)
-\frac{1-\beta^2\tilde{\kappa}^2/\tilde{A}^2\omega_1^2}{z^2_+}\mathbf{K}
\left(1-z^2_-/z^2_+\right)\right], \\ \nn &&\mathcal{J}_2= -\frac{2
z_+ \omega_2/\omega_1 }{\sqrt{1-\omega_2^2/\omega_1^2}}\mathbf{E}
\left(1-z^2_-/z^2_+\right), \\ \nn &&\Delta\varphi_1=
-\frac{2\beta/\tilde{A}}{\sqrt{1-\omega_2^2/\omega_1^2}z_+}
\left[\frac{\tilde{\kappa}^2/\omega_1^2}{1-z^2_+}\Pi\left(-\frac{z^2_+
- z^2_-}{1-z^2_+}|1-z^2_-/z^2_+\right) -\mathbf{K}
\left(1-z^2_-/z^2_+\right)\right].\eea Our next step is to introduce
the new parameters \bea\nn u\equiv\omega^2_2/\omega^2_1,\h v\equiv
\beta/\tilde{A},\h \epsilon\equiv z^2_-/z^2_+ ,\eea and to rewrite
$\mathcal{E}$, $\mathcal{J}_1$, $\mathcal{J}_2$, $\Delta\varphi_1$
in the form (\ref{ncoeffGM}). The explicit $\epsilon$-expansions of
the coefficients $K_e,\ldots,K_{\varphi 3}$ are given as functions
of $u$ and $v$ in appendix A. The $\epsilon$-expansions for $u$ and
$v$ are the same as before. Now, the coefficients in these
expansions can be determined by the condition that $\mathcal{J}_1$
and $\mathcal{J}_2$ should be finite, \bea\nn
&&v_0=\frac{2\mathcal{J}_1}{\sqrt{\left(\mathcal{J}_1^2-\mathcal{J}_2^2\right)
\left[4-\left(\mathcal{J}_1^2-\mathcal{J}_2^2\right)\right]}},\h
u_0=\frac{\mathcal{J}_2^2}{\mathcal{J}_1^2},
\\ \nn
&&v_1=\frac{(1-u_0)v_0^2-1}{4(u_0-1)(v_0^2-1)v_0}
\left\{(u_0-1)v_0^4(1+\log(16))-2\right.\\ \nn &&+ \left.
v_0^2\left[3+\log(16)+u_0(\log(4096)-5)\right]\right\},
\\ \nn
&&v_2=-\frac{v_0\left[1-(1-u_0)v_0^2\right]\left[1+3u_0-(1-u_0)v_0^2\right]}
{4(1-u_0)(v_0^2-1)},\\ \nn
&&u_1=\frac{u_0\left[1-(1-u_0)v_0^2\right]\log(16)}{v_0^2-1},\h
u_2=-\frac{u_0\left[1-(1-u_0)v_0^2\right]}{v_0^2-1}. \eea The
parameter $\epsilon$ can be obtained from $\Delta\varphi_1$ as
follows: \bea\nn \epsilon=16
\exp\left(-\frac{\sqrt{(1-u_0)v_0^2-1}}{v_0^2-1}\left[\Delta\varphi_1
+ \arcsin\left(\frac{2\sqrt{(1-u_0)v_0^2-1}}
{(1-u_0)v_0^2}\right)\right]\right).\eea Taking the above results
into account, $\mathcal{E}-\Delta\varphi_1$ can be derived as
\bea\label{DiffJ12}\mathcal{E}-\Delta\varphi_1&=& \arcsin
N(\mathcal{J}_1,\mathcal{J}_2) +
2\left(\mathcal{J}_1^2-\mathcal{J}_2^2\right)
\sqrt{\frac{4}{\left[4-\left(\mathcal{J}_1^2-\mathcal{J}_2^2\right)\right]}-1}\\
\nn
&\times&\exp\left[-\frac{2\left(\mathcal{J}_1^2-\mathcal{J}_2^2\right)
N(\mathcal{J}_1,\mathcal{J}_2)}
{\left(\mathcal{J}_1^2-\mathcal{J}_2^2\right)^2 +
4\mathcal{J}_2^2}\left[\Delta\varphi_1 +\arcsin
N(\mathcal{J}_1,\mathcal{J}_2)\right]\right],\\
\nn N(\mathcal{J}_1,\mathcal{J}_2)&\equiv&
\frac{1}{2}\left[4-\left(\mathcal{J}_1^2-\mathcal{J}_2^2\right)\right]
\sqrt{\frac{4}{\left[4-\left(\mathcal{J}_1^2-\mathcal{J}_2^2\right)\right]}-1}.\eea
Finally, by using the SS relation between the angular momenta
\bea\nn \mathcal{J}_1=\sqrt{\mathcal{J}_2^2+4\sin^2(p/2)},\eea we
obtain \bea\label{ssS3c} \mathcal{E}-\Delta\varphi_1=
p+8\sin^2\frac{p}{2}\tan\frac{p}{2}
\exp\left(-\frac{\tan\frac{p}{2}(\Delta\varphi_1+ p)}
{\tan^2\frac{p}{2} + \mathcal{J}_2^2 \csc^2p}\right).\eea This is
our final expression for the leading finite-size correction to the
``$E-\Delta\varphi$'' relation for the membrane analogue of the SS
string with two angular momenta. It coincides with the string
result found in \cite{AB1}. As in the GM case, the difference is
in the identification (\ref{not}).

\setcounter{equation}{0}
\section{Concluding Remarks}

In this paper, by using the possibility to reduce the M2-brane
dynamics to the one of the NR integrable system, we gave an explicit
mapping connecting the parameters of {\it all} membrane solutions
described by this dynamical system and the parameters in the
corresponding solutions of the CSG integrable model. Based on this
NR approach, we found finite-size M2-brane solution, its image in
the CSG system, and the leading finite-size corrections to the
energy-charge relations analogous to the ones for the GM and SS
strings on $R_t\times S^3$.

An evident direction for further investigations is to consider more
general membrane configurations, which could describe finite-size
effects corresponding to GM and SS strings on $R_t\times S^5$. i.e.
with three angular momenta. One can also try to include the energy
dependence on the spin $\mathcal{S}$, arising from the $AdS$ part of
the full $AdS_4\times S^7$ background. Note however, that such
general cases are not considered yet even for strings on
$AdS_5\times S^5$.

Another interesting problem is to find the M2-brane analogues of the
semiclassical GM and SS scattering \cite{HM06,CDO06S,IKSV07}. To
this end, one can use the established correspondence between the
membrane solutions (\ref{MGM}), (\ref{MSS}) and the CSG model.
Alternatively, one may apply the dressing method as is done in
\cite{SV06,KSV06,IKSV07}.

\section*{Acknowledgements}
This work was performed at the 2008 APCTP Focus Program ``Finite-size
technology in low-dimensional quantum systems (4)''.
It was supported in part by KRF-2007-313-C00150 (CA), by NSFB
VU-F-201/06 (PB), and by the Brain Pool program from the Korean
Federation of Science and Technology (2007-1822-1-1). PB also
acknowledges the SEENET-MTP support.

\def\theequation{A.\arabic{equation}}
\setcounter{equation}{0}
\begin{appendix}

\section{$\epsilon$-Expansions}

We use the following expansions for the elliptic functions \bea\nn
&&\mathbf{K}(1-\epsilon)\propto
-\frac{1}{2}\log\epsilon\left(1+O(\epsilon)\right)+\log(4)\left(1+O(\epsilon)\right),
\\ \nn
&&\mathbf{E}(1-\epsilon)\propto
1-\epsilon\left(\frac{1}{4}-\log(2)\right)
\left(1+O(\epsilon)\right)-\frac{\epsilon}{4}\log\epsilon\left(1+O(\epsilon)\right)
\\ \nn &&\Pi(n|1-\epsilon)\propto
\frac{\log\epsilon}{2(n-1)}\left(1+O(\epsilon)\right) +
\frac{\sqrt{n}\log\left(\frac{1+\sqrt{n}}{1-\sqrt{n}}\right)-\log(16)}{2(n-1)}
\left(1+O(\epsilon)\right).\eea

The expansions for the coefficients in (\ref{ncoeffGM}) for
$\tilde{A}^2>\beta^2$ are \bea\nn &&K_e \propto
\frac{1-v^2}{\sqrt{1-u-v^2}}-\frac{(1-u)^2-v^2}{2(1-u)
\sqrt{1-u-v^2}}\ \epsilon, \\ \nn &&K_{11}\propto
-\frac{\sqrt{1-u-v^2}}{1-u} -
\frac{\sqrt{1-u-v^2}\left(1-2u-v^2\right)v^2}{2(1-u)^2(1-v^2)}\
\epsilon,
\\ \nn &&K_{12}\propto \frac{(1-u)(1-v^2)}{1-u-v^2}+
\frac{u^2v^2}{(1-u-v^2)(1-v^2)}\ \epsilon,
\\ \nn &&K_{2}\propto
\frac{\sqrt{u(1-u-v^2)}}{1-u} +
\frac{\sqrt{u(1-u-v^2)}\left(1-2u-v^2\right)v^2}{2(1-u)^2(1-v^2)}\
\epsilon,
\\ \nn &&K_{\varphi 1}\propto \frac{v}{\sqrt{1-u-v^2}} - \frac{\left(1-2u-v^2\right)v^3}
{2(1-u)(1-v^2)\sqrt{1-u-v^2}}\ \epsilon,
\\ \nn &&K_{\varphi 2}\propto \frac{1-u}{v^2} -
\frac{u\left(1-u-v^2\right)}{(1-v^2)v^2}\ \epsilon,
\\ \nn &&K_{\varphi 3}\propto 1-\frac{1-u}{v^2} +
\frac{2u\left(1-u-v^2\right)}{(1-v^2)v^2}\ \epsilon .\eea

The expansions for the coefficients in (\ref{ncoeffGM}) for
$\tilde{A}^2<\beta^2$ are given by \bea\nn &&K_e \propto
\frac{v^2-1}{\sqrt{v^2(1-u)-1}}-\frac{v^2(1-u)^2-1}{2
\sqrt{v^2(1-u)-1}(1-u)}\ \epsilon, \\ \nn &&K_{11}\propto
\sqrt{\frac{v^2(1-u)-1}{v^2(1-u)^2}} -
\frac{\sqrt{v^2(1-u)-1}\left(1+v^2(2u-1)\right)}{2v^3(v^2-1)(1-u)^2}\
\epsilon,
\\ \nn &&K_{12}\propto
\left(1-\frac{v^2u}{v^2-1}\right)\epsilon,
\\ \nn &&K_{2}\propto
-\sqrt{\frac{\left(v^2(1-u)-1\right)u}{v^2(1-u)^2}}
+\frac{\sqrt{\frac{\left(v^2(1-u)-1\right)u}{v^2(1-u)^2}}\left(1+v^2(2u-1)\right)}
{2v^2(v^2-1)(1-u)}\ \epsilon,
\\ \nn &&K_{\varphi 1}\propto -\frac{v}{\sqrt{1-1/v^2-u}} - \frac{1+v^2(2u-1)}
{2(v^2-1)\sqrt{v^2(1-u)-1}(1-u)}\ \epsilon,
\\ \nn &&K_{\varphi 2}\propto 1-u +
\frac{\left(1-v^2(1-u)\right)u}{v^2-1}\ \epsilon,
\\ \nn &&K_{\varphi 3}\propto 1-v^2(1-u) + 2v^2u\left(1-\frac{v^2u}{v^2-1}\right)\epsilon
.\eea

\end{appendix}

\end{document}